\documentclass[useAMS,usenatbib,usegraphicx]{mn2e}

\def\hi {\mbox{H\,{\sc i}}} 
\def\hii {\mbox{H\,{\sc ii}}}

\def\kms{km s$^{-1}$}

\title[Neutral hydrogen absorption at the center of NGC\,2146]{Neutral 
hydrogen absorption at the center of NGC\,2146} 
\author[A. Tarchi et al.]{A. Tarchi$^{1,2}$\thanks{E-mail: 
a.tarchi@ira.cnr.it}, A. Greve$^{3}$, A. B. Peck$^{4}$, N. Neininger$^{5}$, K. 
A. Wills$^{6}$, A. Pedlar$^{7}$  
\newauthor and U. Klein$^{5}$\\ 
$^{1}$Instituto di Radioastronomia, CNR, Via Gobetti 101, 40129 Bologna, 
Italy\\ 
$^{2}$Osservatorio Astronomico di Cagliari, Loc. Poggio dei Pini, Strada 54, 09012 Capoterra (CA), Italy\\
$^{3}$IRAM, 300 Rue de la Piscine, F-38406 St. Martin d'H{\`e}res, France\\ 
$^{4}$Submillimeter Array, Harvard-Smithsonian Center for Astrophysics, P.O. 
Box 824, Hilo, HI 96721, USA\\ 
$^{5}$Radioastronomisches Institut der Universit{\"a}t Bonn, Auf dem H{\"u}gel 71, 
D-53121 Bonn, Germany\\ 
$^{6}$Department of Physics and Astronomy, University of Sheffield, Hounsfield 
Road, Sheffield S3 7RH\\ 
$^{7}$Jodrell Bank Observatory, University of Manchester, Macclesfield, Cheshire SK11 9DL, United 
Kingdom} 
 
\begin{document} 
\date{Accepted 1988 December 15. Received 1988 December 14; in original form 
1988 October 11} 
 
\pagerange{\pageref{firstpage}--\pageref{lastpage}} \pubyear{2002} 
 
\maketitle 
 
\label{firstpage} 
 
\begin{abstract}
We present 1.4\,GHz \hi\ absorption line observations towards the starburst 
in NGC\,2146, made with the VLA and MERLIN. The \hi\ gas has a rotating disk/ring structure with column densities between 6 and 18 $\rm \times 10^{21}\:atoms\, cm^{-2}$. The \hi\ absorption has a uniform spatial and velocity distribution, and does not reveal any anomalous material concentration or velocity in the central region of the galaxy which might indicate an encounter with another galaxy or a far--evolved merger. We conclude that the signs of an encounter causing the starburst should be searched for in the outer regions of the galaxy.  
\end{abstract} 
 
\begin{keywords} 
galaxies: starburst -- galaxies: individual: NGC\,2146 -- galaxies: ISM -- 
galaxies: kinematics and dynamics -- radio lines: galaxies. 
\end{keywords} 
 
\section{Introduction} 
Atomic hydrogen absorption measurements on (sub)arc\-second scales allow us to
investigate the distribution and dynamics of neutral gas in the inner regions 
of starburst galaxies and active galactic nuclei (AGN). While \hi\ emission 
measurements with current instruments are sensitivity--limited to a 
maximum resolution of a few arcseconds, absorption studies are limited only by
the angular resolution of the telescope and the brightness distribution of the
background radio continuum. We present \hi\ absorption measurements of the 
peculiar spiral galaxy NGC\,2146. 
 
It is well established that NGC\,2146 is undergoing a strong starburst, even 
stronger than that in M\,82 (e.g.\ {\mbox{\citealt{kronberg81}}}; \citealt{tarchi00}; later TNG); however, the origin of this starburst is still unclear. A starburst is often 
triggered by an interaction with another galaxy, which perturbs the potential 
equilibrium of the gas, causing a flow of gas towards the center which results 
in an increase in density and thus fueling of the star--formation process.

It therefore seems plausible that a starburst in a galaxy is triggered by an encounter if the galaxy belongs to a group which contains and still is connected by large \hi\ tails, for instance like M\,81\,--\,M\,82\,--\,NGC\,3077 (\citealt{yun94}, \citealt{walter02}) and NGC\,3627\,--\,NGC3628\,--\,NGC\,3623 (the Leo triplet; \citealt{zhang93}). Such an interaction has been searched for in NGC\,2146 since the presence of a large \hi\ cloud is known since 1976.
 
Using the NRAO 91--m telescope, \citet{fisher76} detected a huge \hi\ cloud 
extending out to six Holmberg radii ($\sim$\,120\,kpc) around NGC\,2146. This cloud could be the consequence of a tidal interaction or of an
explosion/ejection in the galaxy. However, there exists no kinematic evidence 
for the explosion hypothesis, and no companion has been found which may have 
interacted with NGC\,2146. The higher resolution observations of the extended
\hi\ around NGC\,2146 with the WSRT \citep{caspers86} and the VLA 
\citep{tara96} resolved the cloud into a prominent tail, extending out to 
90\,kpc SE of the body of the galaxy, but no further evidence of an 
interaction was found at that time. In 1990, \citet{hutch90} suggested instead
that NGC\,2146 appears to be in the final stage of a fairly gentle 
far--evolved merger, with the dominant galaxy (NGC\,2146) now seen close to 
edge--on and the stripped companion being on a final plunge toward its center.
The putative traces of this merger are, however, not particularly compelling. 
Evidence of a collision with another galaxy which did not remain embedded in 
NGC\,2146 has been suggested as an alternative triggering mechanism by 
\citet{young88}. They drew especially attention to the 10\,kpc extended 
semi--arc of \hii\ regions (observed in H$\alpha$ and [SII]), which is not 
coplanar with the rotating disk of the galaxy. A more recent analysis by 
\citet{tara01} (later TPB) of the \hi\ stream suggests today a tidal interaction 
between NGC\,2146 and a Low Surface Brightness (LSB) companion of which a 
remnant is apparently still seen as a 1.5$\times$10$^{8}$\,M$_{\sun}$ \hi\ 
concentration in the southern tail. 
 
In order to study the kinematics and density of gas in the central region of 
NGC\,2146 in the light of the proposed merger/encounter hypothesis, we have 
mapped the \hi\ absorption towards the nuclear radio continuum emission using 
the VLA (1\farcs8 resolution = 130\,pc) and MERLIN (0\farcs2 resolution = 
15\,pc). At the distance of NGC\,2146 (14.5\,Mpc, \citealt{benvenuti75}), 
1$''$ is equivalent to 70\,pc. 
 
\section{The observations and image processing} 
 
\textit{\textbf{VLA\,Observation}} The \hi\ line was observed towards 
NGC\,2146 for a total of 11 hours on March 16 (5.5 hrs) and 17 (5.5 hrs), 
2001, with the Very Large Array\footnote{The National Radio Astronomy 
Observatory is a facility of the National Science Foundation operated under 
cooperative agreement by Associated Universities, Inc.} (VLA) in 
A--configuration. We observed with four 6.25~MHz IFs centered on the systemic 
velocity of the galaxy (893 \kms; \citealt{devaucouleurs91}). The QSO 
0542+498 (21.94\,Jy) was used as a flux calibrator; the QSO 0626+820 was used 
for phase and bandpass calibration. The data were Fourier--transformed using 
natural weighting to create a $512\times512\times30$ data cube with a 
restoring beam of $2\farcs1\times1\farcs6$ and a noise level per channel of 
0.2 mJy/beam. The radio continuum was subtracted using the AIPS task UVLSF. 
This task fits a straight line to the real and imaginary parts of selected 
channels and subtracts the fitted baseline from the spectrum, optionally 
flagging data having excess residuals. In addition, this procedure provides 
the fitted continuum as a UV data set, which has been used to create the 
naturally weighted map (Fig.~\ref{contna}). 
 
\noindent\textit{\textbf{MERLIN\,Observation}} The \hi\ line in 
NGC\,2146 was observed in June 1999 for 23 hours with MERLIN (6 antennas). The
total bandwidth of 8 MHz was divided into 32 adjacent channels of 250 kHz 
width (52.6 \kms). Three bad channels, two at the beginning and one at the end
of the band, were removed, so that only 29 channels are present in the 
spectra. The frequency of channel 14 corresponds to a heliocentric velocity of
893 \kms. The re\-lative gains of the antennas were determined using the point 
source calibrator 0552+398, with a flux density of 1.754 Jy. The QSOs 3C286 
(13.7 Jy) and 0602+780 were used for absolute flux density and phase 
calibration, respectively. The data were Fourier-transformed using natural 
weighting to produce a $\mathrm{2048\times2048\times29}$ spectral line data 
cube. A naturally weighted continuum image was obtained by averaging 19 
channels free of line absorption (channels 2 to 10 and 19 to 28). This 
uncleaned image was subtracted from the spectral line cube. Both the continuum 
image and the continuum-subtracted cube were then deconvolved using the CLEAN 
algorithm \citep{hoegbom74} and convolved with the same beam 
($\mathrm{0\farcs20\times0\farcs19}$). The cleaned continuum was added 
back to the continuum-subtracted cube. The rms noise levels in 
source-free areas of the continuum maps and the spectral maps are 
$\mathrm{\sim 65\:\mu Jy\;beam^{-1}}$ and $\mathrm{\sim 300\:\mu 
Jy\;beam^{-1}}$, respectively, consistent with the expected thermal noise. 
 
\section{The results} 
\subsection{Spatial features and sources} 
 
\begin{figure*} 
\centering 
\includegraphics[width=16cm]{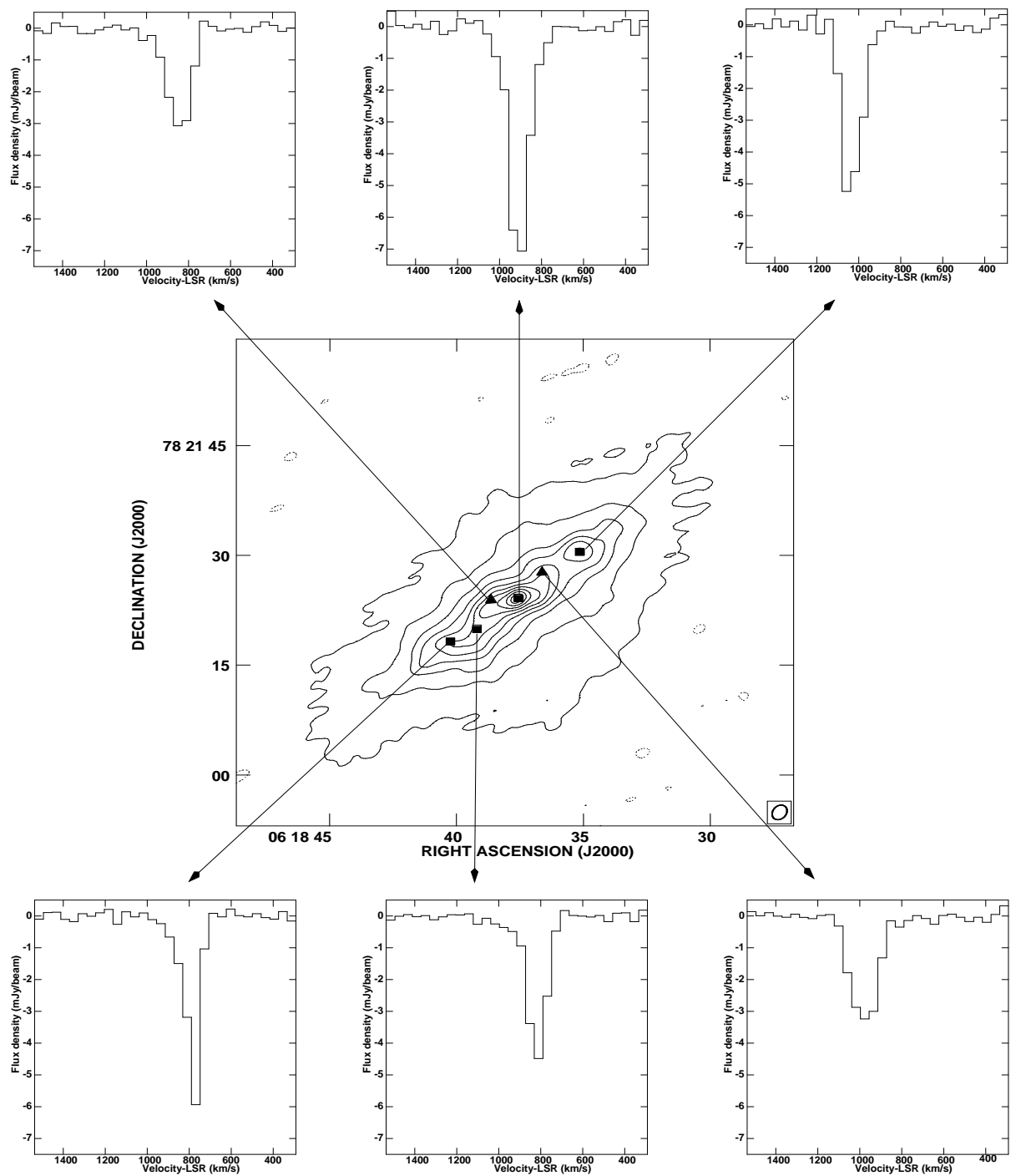}
\caption{Naturally weighted 1.4\,GHz VLA image of NGC~2146 (central panel) and 
\hi\ absorption--line velocity profiles (small panels) obtained at six 
positions: four associated with radio continuum emission peaks (squares) and 
two coincident with the regions containing water masers (triangles). The
angular resolution of the 1.4\,GHz VLA map is $2\farcs1\times1\farcs6$ 
(150\,pc$\times$\,110\,pc); the contour levels are (--\,1, 1, 4, 8, ..., 40) 
$\times$ $6 \times 10^{-4}$ $\rm mJy\;beam^{-1}$. The synthesized beam is 
shown in the lower right corner. Properties of the radio continuum and \hi\ 
absorption at the selected locations are given in Table~\ref{abstabna}.} 
\label{contna} 
\end{figure*} 
 
\begin{figure} 
\centering 
\includegraphics[width=8cm]{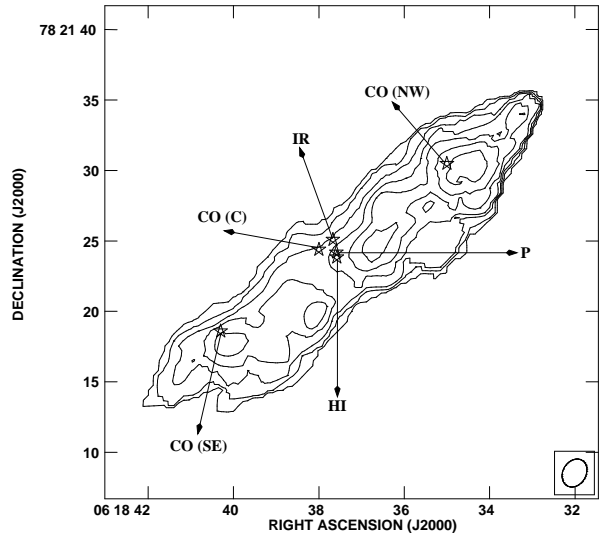} 
\caption{Average \hi\ opacity. The contour increment is 0.02 in optical depth, 
extracted from the naturally weighted VLA optical depth cube integrated over a 
680 $\ldots$ 1160 \kms\ velocity range. The average \hi\ opacity is 
representative of the \hi\ column density. For the inserted symbols see 
Table~\ref{positions}. The synthesized beam is shown in the lower right 
corner.} 
\label{tauna} 
\end{figure} 
  
\begin{figure*} 
\centering 
\includegraphics[width=16cm]{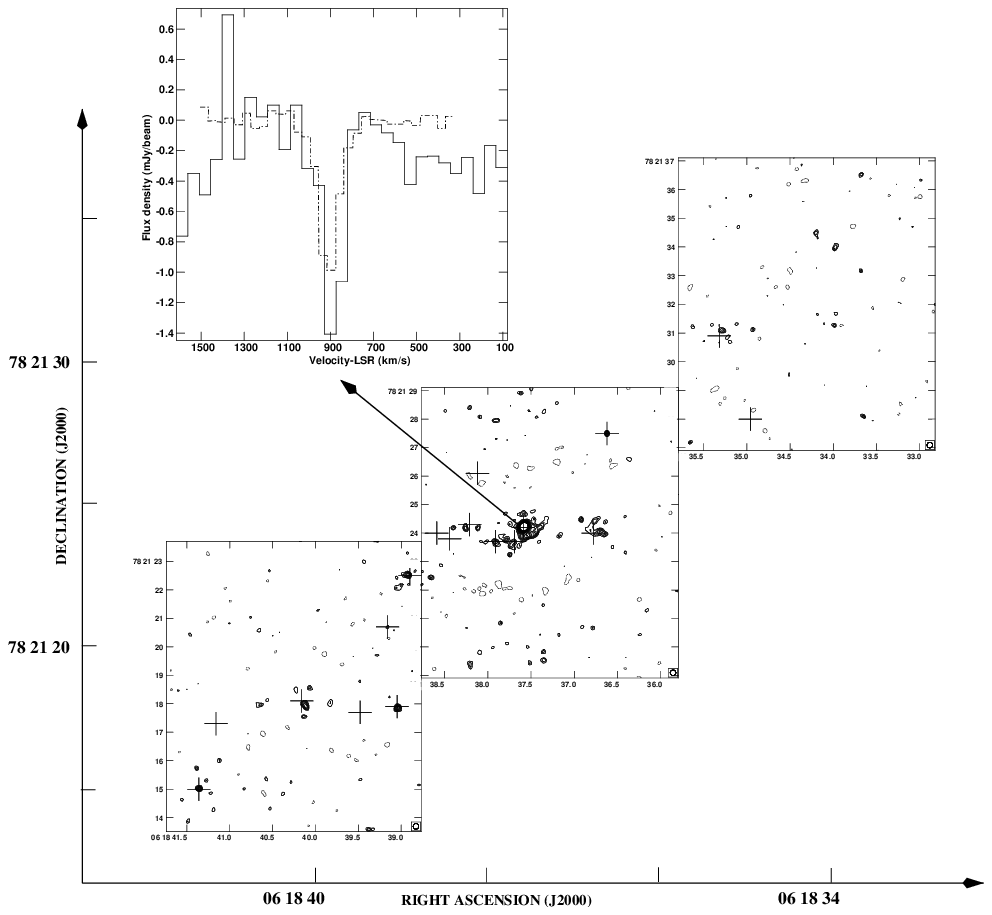} 
\caption{Diagonal panels: A naturally weighted MERLIN radio continuum 
image (resolution $\approx$ $0\farcs2$ $\approx$ 14\,pc) along the major
axis of NGC~2146 at 1.4~GHz, using absorption--free channels. The peak is at 
$\sim$ 4 $\rm mJy\,beam^{-1}$. The contour interval is 0.03 $\rm 
mJy\,beam^{-1}$ with the first contour at 0.13 $\rm mJy\,beam^{-1}$ 
(2$\sigma$). The crosses indicate the positions of the 18 compact sources 
detected in a 5~GHz VLA\,+\,MERLIN map by TNG. The insert shows 
the H\,{\sc i} absorption observed with MERLIN (solid line) and the VLA 
(dot--dash line; the intensity has been multiplied by a factor 1/7) towards 
the central source 37.6\,+\,24.2.} 
\label{absmer} 
\end{figure*} 
 
Fig.~\ref{contna} shows the VLA A--array 1.4\,GHz naturally weighted continuum 
image of the central 4.2\,kpc$\times$1.8\,kpc region of NGC\,2146. The 24\,mJy 
peak is at the center of the galaxy; three other weaker regions of emission 
are also detected. The higher resolution 1.4\,GHz MERLIN map, shown in 
Fig.~\ref{absmer}, resolves these emission peaks into individual 
supernova remnants and compact/ultra--dense \hii\ regions (TNG). 
The inserts in Fig.~\ref{contna} are \hi\ absorption spectra integrated over 
the brightest 9 pixels ($\mathrm{1\arcsec 
\times1\arcsec\approx5\,000\,pc^{2}}$) surrounding six selected regions of 
NGC~2146. Four of these regions are associated with radio continuum emission 
peaks and two coincide with the position of the water masers reported by 
\citet {tarchi02}. 
 
The spectra taken across the 1.4\,GHz map contain the \hi\ absorption line. An 
optical depth cube has been created using the relation 
\begin{equation} 
\mathrm{\tau=-ln(1-T_{L}/T_{C})}, \label{eq:tau} 
\end{equation} 
where $\mathrm{T_{L}}$ is the brightness temperature of the line, 
$\mathrm{T_{C}}$ the brightness temperature of the continuum, and 
$\mathrm{\tau}$ the optical depth of the line. We assume that the adopted 
excitation temperature of $\mathrm {T_{ex}}$ = 100 K is much smaller than 
$\mathrm{T_{C}}$. 
 
Fig.~\ref{tauna} shows the average opacity ($\tau$) distribution derived in 
this way from the optical depth cube, blanking pixels with signal--to--noise 
ratio below 4. This map is not biased by the strength of the continuum. The 
\hi\ absorption is seen as an elongated region of 2.4\,kpc$\times$0.6\,kpc 
extent, thus similar in extent to the starburst region, at a position angle of
the major axis of NGC\,2146 ($\mathrm PA \approx $ 140$^{\rm o}$). The extent 
and axial ratio of the \hi\ absorption (and of the underlying emission) region
of $\sim$\,1/4 is also seen in synchrotron emission maps taken at higher 
frequencies by \citet{lisenfeld96}. Superimposed on the map of 
Fig.~\ref{tauna} are the positions of some prominent features in the 
nuclear region of NGC~2146, viz.\ the \hi\ kinematic center (HI), the CO 
kinematic center (CO(C)), the 2.2\,$\mu$m peak of stellar radiation (IR), and 
the strong central radio continuum point source (P). For the position of the 
\hi\ kinematic center we adopt the \hi\ emission peak at the systemic velocity
(see Fig.~\ref{kntrna}); the CO kinematic center is taken from \citet{greve04}; the 2.2\,$\mu$ peak is taken from the 2MASS Extended Source Image 
Server of the NASA/IPAC Infrared Science Archive 
(http://irsa.ipac.caltech.edu/index.html); the position of the central source 
is taken from TNG. The coordinates of these features are given in
Table~\ref {positions}. When we consider that the CO emission peaks have extensions of a few hundred parsecs, we find from Fig.\,2 that the \hi\ opacity peaks coincide clearly with the tangential directions CO(NW,\,SE) of the molecular ring, but also to some extent with the central CO emission. It 
appears that the kinematic center of the \hi\ gas coincides with the center of
the molecular gas and the stars of the starburst region (see Fig.~\ref{tauna} 
in \citealt{young88}) to within the errors of the measurements ($\pm$ 
100\,pc). Since in addition the orientation (PA) of the \hi\ gas agrees with 
the orientation of the molecular gas and the stars, we believe that the \hi\ 
gas is to a large extent co--spatial with these other constituents, thus 
belonging to the same kinematic body of material of the starburst region.   
   
\begin{table*} 
\caption{Positions of important features in the nuclear region of NGC~2146.} 
\label{positions} 
\begin{tabular}{llll} 
\hline 
Label  & Description  & R.A. (2000) & Dec. (2000)      \\ 
    &        & 06$\mathrm{^{h}}$18$\mathrm{^{m}}$ & 78\degr21\arcmin \\ 
\hline 
HI    & \hi\ kinematic center  & 37\fs58  & 23\farcs85    \\ 
CO(SE)   & SE peak CO ring         & 40\fs3   & 18\farcs6     \\
CO(C)   & CO central peak        & 38\fs0   & 24\farcs4     \\ 
CO(NW)   & NW peak CO ring        & 35\fs0   & 30\farcs5     \\ 
IR   & 2.2 $\rm \mu m$ peak (stars) & 37\fs67 & 25\farcs1     \\ 
P    & Central radio point source  & 37\fs59  & 24\farcs2   \\ 
\hline 
\end{tabular} 
\end{table*} 
 
We have used the optical depth cube to compute the peak and the integrated 
optical depth of the lines extracted at the six positions mentioned above. The
column density N (in $\mathrm{atoms\:cm^{-2}}$) of the absorbing material is 
calculated from 
\begin{equation} 
\mathrm{N= \frac{T_{ex}}{5.49\times10^{-19}} \int \tau \it{d} v}, 
\label{eq:density} 
\end{equation} 
where $\mathrm{\it{d} v}$ = 52.6 $\mathrm{km\:s^{-1}}$ is the velocity 
resolution of the observation, and $\mathrm{T_{ex}}$ = 100\,K as typically 
used for starburst galaxies (e.g.\ \citealt{welia84}). The quantities derived 
in this way are listed in Table~\ref{abstabna}. Columns~1 and 2 give the radio
continuum peak locations at which the spectra of Fig.~\ref{contna} are taken; 
their 1.4\,GHz continuum peak flux densities are listed in Column~3. Columns~4
to 8 summarize the \hi\ absorption properties at these locations, i.e.\ the 
line velocity and width, and the peak and integrated optical depth derived 
from Gaussian profiles fitted to the spectra. Column~9 gives the corresponding
\hi\ column density. 
 
\begin{table*} 
\caption{Representative \hi\ absorption at six locations within the central 
region of NGC\,2146, extracted from the naturally weighted map made with the 
VLA. S$\mathrm{_{1.4}}$: 1.4\,GHz flux density;. {\it v} and 
$\mathrm{\Delta}${\it v}: line velocity and half--power line width 
[the systemic velocity is 893 $\mathrm{km\:s^{-1}}$ ({\it v}$_{lsr}$)]. The 
errors are given in brackets. Boldface: the \hi\ absorption feature detected 
with MERLIN.} 
\label{abstabna} 
\begin{tabular}{ccccccccc} 
\hline 
R.A.       &   Dec. & S$\mathrm{_{1.4}}$ & $v_{\hi}$ & $\Delta {\it v}_{\hi}$ & $\mathrm{\tau_{peak}}$ & $\mathrm{\int{\tau dv}}$ & N & direction \\ 
06$\mathrm{^{h}}$18$\mathrm{^{m}}$ & 78\degr21\arcmin & $\rm (mJy\:beam^{-1})$ & (\kms) & (\kms) &  & & $\mathrm{\times10^{21}\:atoms\:cm^{-2}}$ & of \\ 
(J2000) & (J2000) & & & (FWHM) & & & \\ 
\hline 
35\fs15 & 30\farcs5 & 12.2 (0.2) & 1035 (3) & 100 (7) & 0.60 (0.03) & 64 (4) & 11.7 (0.7) & \\ 
36\fs64 & 27\farcs7 & 11.3 (0.2) & 978(4) & 136 (8) & 0.37 (0.02) & 53 (3) & 9.7 (0.5) & H$_{2}$O maser \\ 
37\fs58 & 24\farcs2 & 23.7 (0.2) & 907 (1) & 99 (3) & 0.40 (0.01) & 42 (1) & 7.7 (0.2) & central source \\ 
 {\bf 37\fs58} & {\bf 24\farcs2} & {\bf 3.99 (0.08)}& {\bf 879  (6)} & {\bf 99 (16)} & {\bf 0.48 (0.1)} & {\bf 48 (5)} & {\bf 9.0 (1)} & central source \\ 
 38\fs63 & 24\farcs0 &  15.1 (0.2) & 842 (2) & 115 (5) & 0.25 (0.02) & 30 (1) & 5.5 (0.2) & H$_{2}$O maser \\ 
 39\fs20 & 20\farcs0 &  12.7 (0.2) & 817 (1) & 91 (3) & 0.46 (0.03) & 44 (1) & 8.0 (0.2) & \\ 
 40\fs24 & 18\farcs2 &  12.0 (0.2) & 778 (0.1) & 58 (0.2) & 0.75 (0.04) & 47 (0.1) & 8.6 (0.02) & \\ 
\hline 
\end{tabular} 
\end{table*} 
 
In the 1.4\,GHz naturally weighted continuum MERLIN map (Fig.~\ref{absmer}) we 
find 10 of the 18 compact sources detected at 5\,GHz as reported by 
TNG. Of these 10 sources, only the central one, 37.6+24.2 (for 
the source notation see TNG), shows a clear \hi\ absorption 
line as illustrated in Fig.~\ref{absmer}. This MERLIN spectrum is obtained 
from an integration over the brightest 9 pixels surrounding the source peak 
($\mathrm{0\farcs045\times0\farcs045\approx10\:pc^{2}}$). The absorption 
properties of this line are derived in the same way as described for the VLA 
data and are reported in boldface in Table~\ref{abstabna}; there exists good 
agreement between the VLA and MERLIN data. The higher sensitivity VLA 
observation does not confirm the three other weak absorption features 
tentatively detected with MERLIN and reported by \citet{tarchi01}. The other 
1.4\,GHz MERLIN detected sources are too weak at the MERLIN resolution, and do
not show any absorption above the noise level. When assuming an excitation temperature of 100\,K and a linewidth of $\sim$ 100 \kms (two channels), the indicative upper limit for a compact source of 0.7 $\rm mJy\:beam^{-1}$ is $3.5\times10^{22}$ $\rm atoms\:cm^{-2}$. This limit is consistent with the \hi\ column density derived from the VLA data.
 
\subsection{Rotation velocities} 
 
\begin{figure*} 
\centering 
\includegraphics[width=16cm]{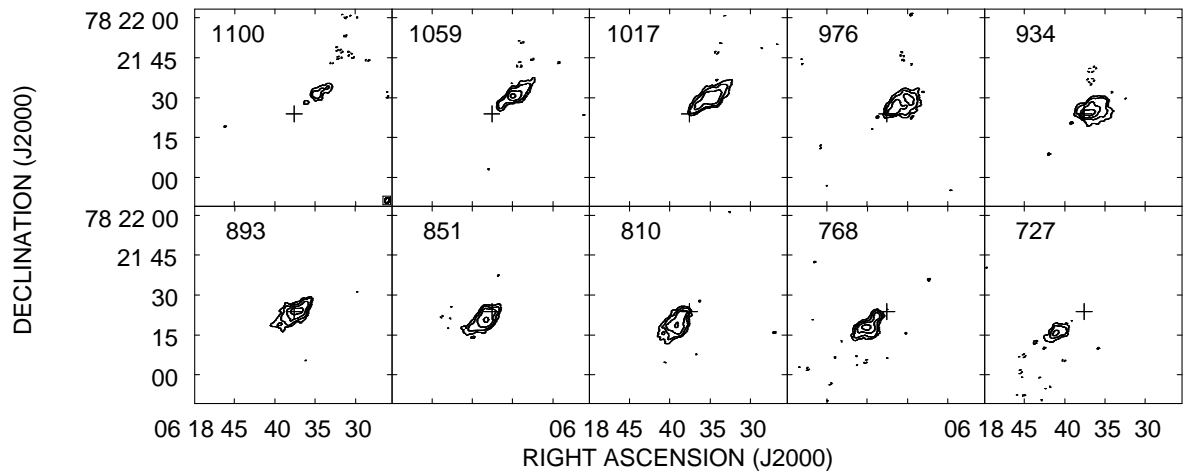} 
\caption{Velocity channel maps extracted from the naturally weighted, 
continuum-subtracted spectral line cube, for 10 velocities (given in \kms\ in 
the upper--left corner) over which the \hi\ absorption is present. The cross 
indicates the center of the absorption at the systemic velocity (893 \kms). 
The absorption is shown as solid contours of (--\,10, 10, 25, 50, 100) 
$\times$ $10^{-4}$ $\rm mJy\;beam^{-1}$.} 
\label{kntrna} 
\end{figure*}

\begin{figure} 
\centering 
\includegraphics[width=8cm]{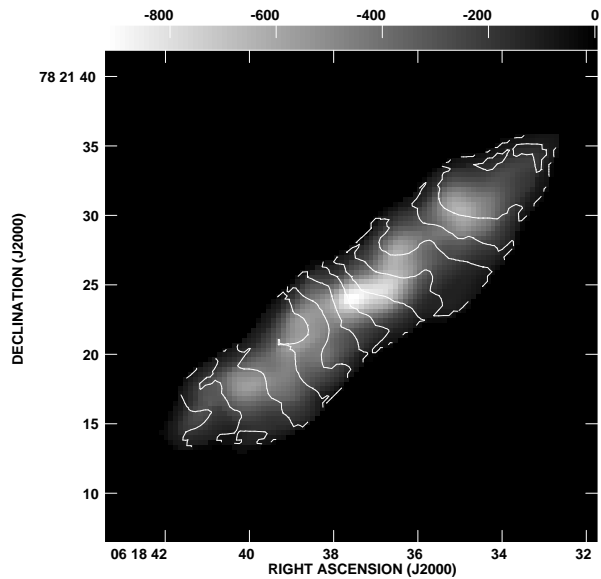} 
\caption{Integrated absorption extracted from the naturally weighted continuum 
subtracted spectral line cube, with the grey--scale flux density from --\,920 
to 0 $ \rm mJy\,beam^{-1}\,km\,s^{-1}$. The absorption shown in this way is 
dependent on the strength of the continuum emission. Superimposed are the \hi\
velocities, with contours from 740 to 1060 \kms, in steps of 25 \kms.} 
\label{mom01na} 
\end{figure}

Individual velocity channel maps with significant \hi\ absorption extracted 
from the continuum--subtracted naturally weighted cube are shown in 
Fig.~\ref{kntrna}. Absorption can be traced over 10 channels, with the maximum
of the emission shifting with increasing velocity from the east 
($\sim$\,1\,kpc) to the west ($\sim$\,0.9\,kpc) of the center. This figure 
clearly shows the rotation of the \hi\ absorption material, and also indicates
the feature of a ring--like emission as seen especially in the CO observations
(\citealt {jackson88}; \citealt{young88}). Fig.~\ref{mom01na} shows the 
integrated total intensity map of the \hi\ absorption, extracted from the 
continuum subtracted cube, with superimposed the isovelocity contours of the 
absorbing \hi\ gas. Using a systemic velocity of 893\,km\,s$^{-1}$, the rotation of the disk of \hi\ gas increases from --\,150\,km\,s$^{-1}$ in the east to +\,170\,km\,s$^{-1}$ in the west.
 
\begin{figure*} 
\centering 
\includegraphics[width=16cm]{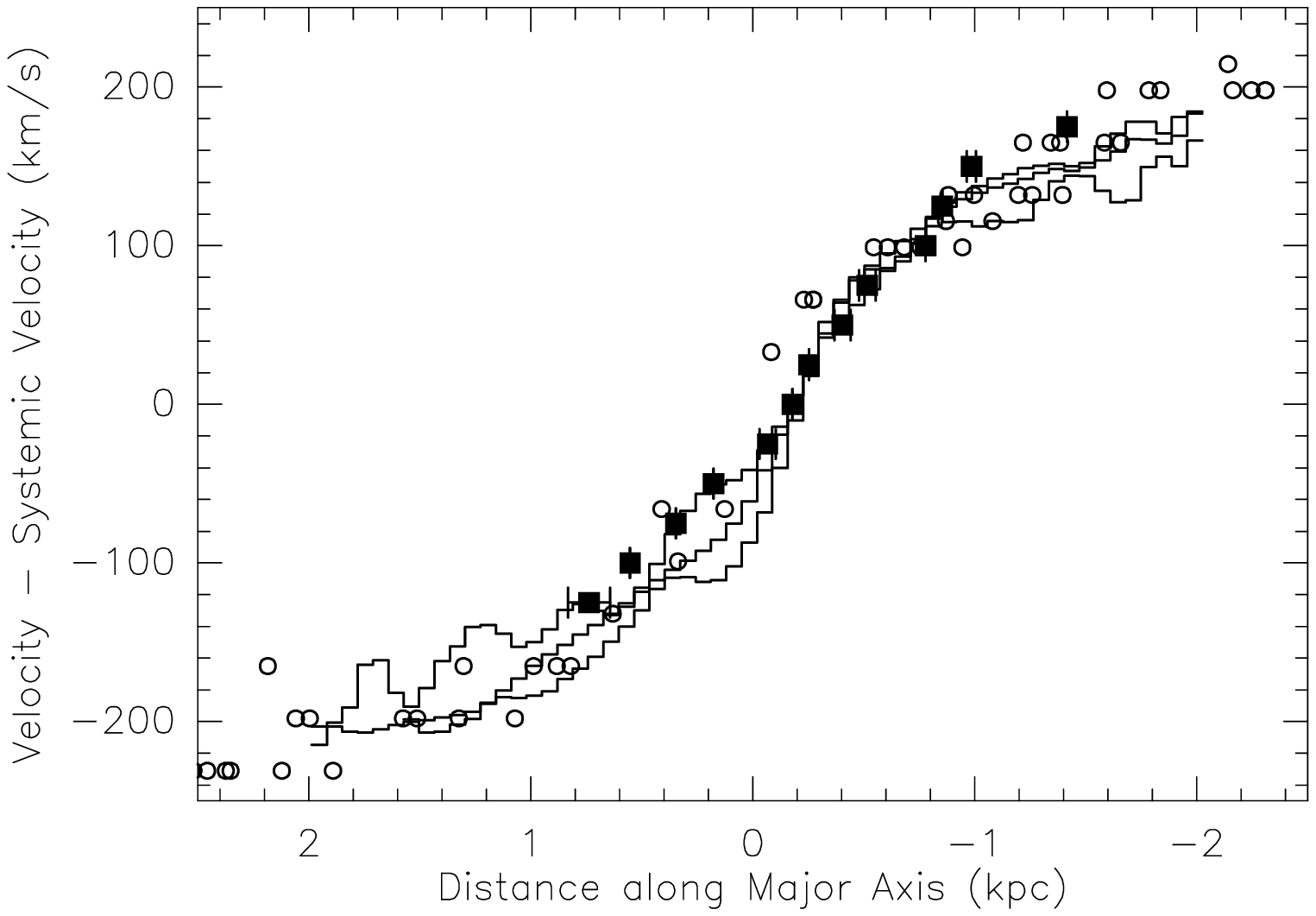} 
\caption{Position--velocity diagrams taken along the major axis of NGC~2146. 
Histogram lines: $^{12}$CO(1-0), (2--1) and $^{13}$CO(1--0); open circles: 
H$\alpha$ from \citet{benvenuti75}; filled squares: \hi\ absorption.
[East is to the left, West to the right].} 
\label{pvothers} 
\end{figure*}

Finally, in Fig.~\ref{pvothers} we compare the rotation velocity of the \hi\ 
absorption with the rotation velocity of the molecular gas (CO) and the 
ionized gas (H$\alpha$). We find that the \hi\ material rotates in the same 
way as the other constituents of the starburst region. This fact confirms that
the \hi\ gas seen in absorption is concentrated in, or close to, the starburst
region. In addition we find that the rotation of the \hi\ gas, the molecular 
gas, the ionized gas, and the stars within $\sim$\,$\pm$\,2\,kpc of the 
starburst region (\citealt{greve04}) is smooth and does not reveal evidence 
of any disturbance from a galaxy encounter or a merger.  
 
The \hi\ line width (Table~\ref{abstabna}) agrees well with the width of the 
CO lines and the width of emission lines originating in the ionized gas. The 
spectra of Fig.~\ref{contna} and the values in Table~\ref{abstabna} suggest 
that the line widths of the absorption features at the maser positions are 
systematically broader by 28 $\pm$ 5 \kms\ than those at the 
other closeby positions, perhaps indicating a more turbulent medium
 in the recent star formation region.
 
\section{Discussion} 
The only previous single--dish study of \hi\ gas in NGC~2146 by 
\citet{fisher76} did not reveal any absorption, because of the low 
spatial resolution of the observation ($\sim$ 10\arcsec = 700\,pc). The 
first detection of \hi\ absorption in NGC~2146 is mentioned by \citet{tara96}; later TPB reported details of the 
detection. They found that the absorption is lying in the region of the 
obscuring dust, having a velocity width of $\sim$ 350 \kms\ (thus 
covering the whole velocity range of rotation) and an average optical depth 
of 0.03. However, their VLA \mbox{D--array} observation was unable to spatially resolve the absorption. The higher spatial resolution VLA 
\mbox{A--array} observation presented here reveals \hi\ absorption in front 
of the radio continuum emitted in the central {\mbox{$\sim$\,2\,kpc} of 
NGC~2146, allowing the construction of a map of its spatial distribution and 
its rotation. When discussing \hi\ absorption measurements it must be 
remembered that any absorption depends on the continuum strength against which
it is seen. The optical depth images, on the other hand, are less dependent, 
if not independent, of the continuum and represent the column density 
distribution of \hi\ (Eq.(2)). The strongest absorption is therefore observed 
against the brightest compact radio source at the center of NGC~2146, while in
the optical depth map of Fig.~\ref{tauna} the regions with the highest column 
density of \hi\ are instead distributed almost symmetrically with respect to 
the central source. The optical depth lies between a minimum value of 0.3 and 
a maximum of 0.9, which for a typical linewidth of 100 \kms\ corresponds to 
\hi\ column densities between 6 and 18 $\rm \times 10^{21}\:atoms\, cm^{-2}$, 
respectively. When using the relation between the \hi\ column density $\rm 
N_{\small \hi}$ (cm$^{-2}$) and the optical extinction $\rm A_{V}({\rm mag}) =
 0.62 \cdot 10^{-21}\cdot \:N_{\small \hi}$ derived by \citet{savage79}, and 
assuming for NGC\,2146 the same gas--to--dust ratio as in our Galaxy, we 
obtain $\rm A_{V}$ = 4 $\ldots$ 11$^{\rm m}$ which is consistent with 
the extinction of 4$^{\rm m}$ $\ldots$ 7$^{\rm m}$ derived by e.g.\ \citet{benvenuti75}, \citet{smith95}, and \citet{jackson88}. 
 
The \hi\ absorption velocity field (Fig.~\ref{mom01na}) is very smooth and 
similar to the rotation of the molecular and ionized gas as shown in 
Fig.~\ref{pvothers}, which supports the structure of a rotating disk of \hi\ 
gas. The coincidence of the spatial distribution of the \hi\ material with 
the extent of the continuum radiation of the starburst region, and the 
agreement with the rotation curves of the other constituents, 
suggests that the observed \hi\ material is close to, or mixed with, the 
material of the central region. This is also supported by the fact that the \hi\ absorption seems to trace the molecular ring (see Fig.\,2) although it is not seen as a distinct feature in the synchrotron radiation (\citealt{lisenfeld96}). The rotation of the gas shown in Fig.~\ref{pvothers}
indicates a dynamic mass inside 2\,kpc of 1.2$\times$10$^{10}$\,M$_{\sun}$, 
for a galaxy inclination between 65\degr to 60\degr. The dynamic mass 
concentrated in this region is probably about 1/5\,th, or less, of the total 
mass of the galaxy of at least 5$\times$10$^{10}$\,M$_{\sun}$ (Benvenuti et 
al.\ 1975). 

Neutral hydrogen absorption studies of other starburst (and Seyfert) galaxies appear to trace gas in the central disks of the host galaxies to radial extents of a few hundred parsec only (\citealt{gallimore99}). The \hi\ gas is usually aligned to the larger outer disk of the host galaxy. None of the \hi\ absorption studies of the other starburst galaxies shows a clear indication of an interaction by infall or outflow of gaseous material, and NGC\,2146 is in this respect no exception. Following the argumentation of Gallimore et al. (1999, their Eq.(2)), the {\mbox{parameters}} N$_{21}$ $\approx$ 5$\times$10$^{21}$\,cm$^{-2}$ (Table 2), {\it v}$_{\rm rot}$ $\approx$ 150\,km\,s$^{-1}$ at {\rm r} $\approx$ 1\,000\,pc (Fig.\,6) give a surface density ratio of the visible disk of \hi\ absorption material of $\sum_{\rm HI} / \sum_{\rm C} = 0.028 N_{21} v_{\rm rot} / r \approx 0.02$ which is well below the critical density necessary to push the disk to instability. 
 
In other starburst galaxies the interaction is primarily evident from extended \hi\ emission tails linking the interacting companions, as for example clearly seen in the M\,81--tiplet (\citealt{yun94}, \citealt{walter02}) and the NGC\,3628--triplet (\citealt{schmelz87}). This interaction has 
apparently produced a bar in M\,82 and NGC\,3628, which fuels the starburst. 
While NGC\,2146 has prominent \hi\ tails, a link to an interacting companion 
has not been found, and a bar has not been established.

While numerical simulations do confirm that mergers and flyby encounters (with proper geometrical constraints) can be successfull triggers of starburst (e.\ g.\ \citealt{mihos92}), the rapid gas depletion due to the star formation activity limits the duration of such starbursts to between 50 to 150 Myr depending on the interaction's details (e.\ g.\ \citealt{mihos96}). 

For NGC~2146, TPB conclude that the trajectory of extended \hi\ gas suggests that a possible interaction was in its most violent phase about $10^{9}$ years ago. Then there should have been enough time for the gas in the inner region to relax, as observed, into a disk\footnote{A simplistic upper limit for the relaxation time ($\tau_{\rm rx}$) can be obtained from the time of a perturbing wave, caused by the interaction, to cross, at a velocity ($v$) equal to the inner-galactic sound speed ($\sim 100$ \kms), the starburst region with a diameter ($D$) = 3 kpc. This time is $\tau_{\rm rx} = D / v = 10^{8}$ years}. However, the starburst, if coeval with the interaction, should be already terminated. 
On the contrary, the starburst in NGC~2146 is presently powerful (e.g.\ \citealt{kronberg81}) and relatively young (e.g.\ TNG), e.\ g.\ younger than the one in M~82, and thus contradicting a possible age of 10$^{9}$ years. 
The additional suggestion by TPB that the extended gas is falling into, and will continue to fall into NGC~2146 could explain how the present starburst is fueled. However, in this case we would also expect a sign of gas inflow, while we find instead that the distribution of the gas and stars in the inner $\sim$\,6\,kpc is undisturbed and smooth. 

We might speculate that recursive short ($\sim$ $10^{7}$ years) starbursts have occured since the merging/tidal encounter event, and that today we are witnessing the youngest of them, as it has likely happened in M~82 (e.\ g.\ de Grijs, O'Connell \& Gallagher 2001). In such a case the necessary amount of gas to fuel the star formation activity would be smaller and the infall of gas less mandatory. This speculation does, however, not cast any light on the origin of the starburst.
 
Whatever possibility may be correct, the origin and fueling mechanism of the starburst in NGC~2146 must, in our opinion, be searched for at large distances from the center, i.e. outside the body of the galaxy, where distortions of the distribution of gas and stars are seen.

\section{Summary and Conclusions}

From the 1.4\,GHz \hi\ absorption line observations towards the starburst
in NGC\,2146, made with the VLA and MERLIN, we reach the following conclusions:

\noindent{-- the \hi\ absorption is observed over the entire continuum radiation background which coincides with the starburst region;}

\noindent{-- the \hi\ gas towards the starburst region has a column density between 6 and 18 $\rm \times 10^{21}\:atoms\:cm^{-2}$;}

\noindent{-- the \hi\ velocity field has the structure of a rotating disk and ring, as often found in the inner region of similar starburst systems;}

\noindent{-- the \hi\ absorption has a uniform spatial and velocity distribution without any indication of an encounter with another galaxy or a far-evolved merger; there is no indication of a bar;}

\noindent{-- the absence of anomalous gas concentration or velocities and the agreement of the \hi\ absorption rotation curve with those of different constituents (CO and H$\alpha$), suggest that the inner gas has had time to relax after the encounter; the signs of an encounter causing the starburst must be searched for in the outer regions of the galaxy.}  

\section{Acknowledgments} 

We are grateful to the VLA and MERLIN staff for technical support. A.T. would 
like to thank Daniela Vergani and Filippo Fraternali for insightful 
discussions, and Tom Muxlow, Simon Garrington, and Matteo Murgia for useful comments. We thank the referee for his/her pointed and clarifying comments. This research has made use of the NASA/IPAC Infrared Science Archive, which is operated by the Jet Propulsion Laboratory, California Institute of Technology, under contract with the National Aeronautics and Space Administration.

\bsp 
 
\label{lastpage} 
 
\end{document}